\documentclass[aps,prl,twocolumn,showpacs,superscriptaddress,nofootinbib,floatfix]{revtex4}

\usepackage{graphicx,epic,eepic}
\usepackage{epsfig}
\usepackage{amsmath}

\newcommand{\be}{\begin{eqnarray}}
\newcommand{\ee}{\end{eqnarray}}
\newcommand{\bea}{\begin{eqnarray}}
\newcommand{\eea}{\end{eqnarray}}

\newcommand{\ket}[1]{\mbox{$| #1 \rangle$}}
\newcommand{\braket}[2]{\mbox{$\langle #1  | #2 \rangle$}}
\newcommand{\proj}[1]{\mbox{$|#1\rangle \!\langle #1 |$}}

\def\ch{\raisebox{0.3ex}{$\chi$}}
\def\H{{\cal H}}

\def\O{{\cal O}}

\DeclareMathOperator{\tr}{tr}

\begin{document}


\title{Efficient simulation of one-dimensional quantum many-body systems}

\author{Guifr\'e Vidal}
\affiliation{Institute for Quantum Information, California Institute of
             Technology, Pasadena, CA 91125, USA}

\date{\today}

\begin{abstract}
We present a numerical method to simulate the time evolution, according to a Hamiltonian made of local interactions, of quantum spin chains and systems alike. The efficiency of the scheme depends on the amount of the entanglement involved in the simulated evolution. Numerical analysis indicate that this method can be used, for instance, to efficiently compute time-dependent properties of low-energy dynamics of sufficiently regular but otherwise arbitrary one-dimensional quantum many-body systems.
\end{abstract}


\maketitle


Since the early days of Quantum Theory, progress in understanding the physics of quantum many-body systems has been hindered by a serious, well-known computational obstacle. The number of parameters required to describe an arbitrary state of $n$ quantum systems grows {\em exponentially} with $n$, a fact that renders the simulation of generic quantum many-body dynamics intractable. 

A number of techniques have been developed to analyze specific quantum many-body problems. Exactly and quasi-exactly solvable models \cite{exact} offer valuable insight in particular cases, and their solutions can be used as the basis for perturbative studies. With quantum Monte Carlo (QMC) calculations \cite{QMC}, properties of the ground state of a large class of many-body Hamiltonians can be evaluated. In one dimensional settings, including quantum spin chains, ground-state expectation values for local observables, such as the energy and two-point correlation functions, can be computed with extraordinary accuracy with the density matrix renormalization group (DMRG) \cite{DMRG}. 

The importance and ingenuity of these methods cannot be understated. However, solvable models and their extensions using perturbation theory apply to a very restricted class of physical systems, whereas QMC and DMRG calculations produce reliable outcomes mainly only for static properties of certain ground states. In particular, the efficient simulation of time-dependent properties remains an open problem in most non-perturbative cases. In addition to severely limiting our understanding of quantum collective phenomena, such as high-$T_C$ superconductivity and quantum phase transitions, the inability to efficiently simulate quantum dynamics has deep implications in Quantum Information Science \cite{book}. It is also a practical obtacle for the development of technology based on engineered quantum systems.

In this paper we describe a numerical scheme for simulating certain quantum many-body dynamics. We explain how to efficiently simulate Hamiltonian evolutions in one-dimensional arrays of quantum systems, such as quantum spin chains, with a computational cost that grows {\em linearly} in the number $n$ of systems. The key idea is to exploit the fact that in one spatial dimension, low energy quantum dynamics are often only slightly entangled. We employ a technique developed in the context of quantum computation \cite{efficient} ---thereby illustrating how tools from Quantum Information Science may find applications in other areas of Quantum Physics (see also \cite{Pres,trans,Nielsen}). The present numerical method can be used, for instance, to determine the time-dependent vacuum expectation value 
\be
\langle O^{\dagger}_{x,t}O^{}_{0,0}\rangle,
\label{eq:expect}
\ee 
where $O$ is a local Heisenberg operator. In some cases, and for specific choices of $O$, the Fourier transform $S(k,\omega)$ of (\ref{eq:expect}) is accessible experimentally, e.g. through neutron scattering for phonon dynamics in a solid. This allows for a direct comparison of experimental and simulated data.


We consider a one-dimensional array of quantum systems, labeled by index $l$, $l\in\{1, \cdots, n\}$, each one described by a local Hilbert space $\H_d$ of finite dimension $d$. We assume a collective evolution of the $n$ systems according to a (possibly time-dependent) Hamiltonian $H_n$ made of {\em local} interactions, i.e. $H_n$ is a sum of terms each one involving at most $k$ systems, for a small $k$ independent of $n$. Physically reasonable Hamiltonians are of this form. Given a pure state $\ket{\Psi} \in {\H_d}^{\otimes n}$ of the $n$ systems, the entanglement between a block $A$ containing the $m$ first systems, $l\in \{1,\cdots, m\}$, and a block $B$ containing the $n\!-\!m$ remaining systems can be characterized by the rank $\ch_{A}$ of the reduced density matrix $\rho_A$ for block $A$,
\be
\ch_A \equiv \mbox{rank}(\rho_A), ~~~~~~~~ \rho_A \equiv \tr_B(\proj{\Psi}).
\ee 
Central in the present discussion is parameter $\ch$,
\be
\ch \equiv \max_{m} \ch_A,
\label{eq:max}
\ee
the maximum, over the size $m\in\{ 1,\cdots, n\!-\!1\}$ of block A, of the entanglement between blocks $A$ and $B$ \cite{compare}. 

The {\em essential} requirement for the simulation to be efficient is that the entanglement $\ch$ (in practice a related, effective parameter $\ch_{\epsilon}$) remains ``small'' during the simulated dynamics, in a sense to be specified later on. Numerical tests indicate that this condition is usually met, for instance, when the Hamiltonian $H_n$ contains only short-range interactions and, in particular, in dynamics involving the ground state of $H_n$ and its low-energy excitations. Below we assume that $H_n$ contains precisely only short-range interactions in an {\em open} chain, since these additional, {\em non-essential} requirements significantly simplify the discussion. More specifically, we consider a Hamiltonian made of arbitrary single-body and two-body terms, with the interactions restricted to nearest neighbors, 
\be
H_n = \sum_{l=1}^n K_1^{[l]} + \sum_{l=1}^{n-1} K_2^{[l,l+1]}.
\label{eq:Hamiltonian}
\ee

The simulation scheme is based on an efficient decomposition for {\em slightly entangled} states of $n$-systems and on a protocol to efficiently update this decomposition when a unitary transformation is applied to one or two (nearest neighbor) systems, as described next.


{\em Efficient decomposition.---} State $\ket{\Psi} \in {\H_d}^{\otimes n}$ can be expressed in terms of ${\cal O}(nd\ch^2)$ parameters by first expanding it in a product basis,
\be
\ket{\Psi} = \sum_{i_1=1}^d\!\cdots \!\sum_{i_n=1}^d c_{i_1 \cdots i_n} ~\ket{i_1}\otimes \cdots \otimes \ket{i_n},
\label{eq:compdeco}
\ee
and by then writing the $d^n$ complex coefficients $c_{i_1 \cdots i_n}$ in terms of $n$ tensors $\Gamma^{[l]}$ and $n\!-\!1$ vectors $\lambda^{[l]}$,
\bea
c_{i_1i_2 \cdots i_n} = \!\!\! \sum_{\alpha_1,\cdots,\alpha_{n\!-\!1}}\!\!\! \Gamma_{\alpha_1}^{[1]i_1}  \lambda^{[1]}_{\alpha_1} \Gamma_{\alpha_1 \alpha_2}^{[2]i_2}  \lambda^{[2]}_{\alpha_2} \Gamma_{\alpha_2 \alpha_3}^{[3]i_3}  \cdots  \Gamma_{\alpha_{n\!-\!1}}^{[n]i_n},
\label{eq:superdeco}
\eea
where each $\alpha$ runs from 1 to $\chi$. We refer to \cite{efficient} for a detailed explanation on how to obtain this decomposition, denoted ${\cal D}$ from now on.

When restricted to translationally invariant states in one spatial dimension, decomposition ${\cal D}$ has two main precedents. Except for a number of technical details, it corresponds to a construction introduced by Fannes, Nachtergaele and Werner to study the so-called {\em finitely correlated states} \cite{Fannes} ---in turn a generalization of {\em valence bond states} as analyzed by Affleck, Kennedy, Lieb and Tasaki \cite{Affleck}--- and to \"Ostlund and Rommer's description of {\em matrix product states} \cite{Oestlund} used to analytically study the fixed points of the DMRG method. Here we will not require translational invariance and, most importantly, we will consider the time-dependent case.
 
Vector $\lambda^{[l]}$ in ${\cal D}$ contains the decreasingly ordered coefficients $\lambda^{[l]}_1 \geq \lambda^{[l]}_2 \geq \cdots \geq \lambda^{[l]}_{\chi} \geq 0$ of the Schmidt decomposition \cite{Schmidt} of $\ket{\Psi}$ according to the splitting $A:B$, where $A=[1, \cdots, l]$,
\be
\ket{\Psi} = \sum_{\alpha=1}^{\ch} \lambda^{[l]}_{\alpha} \ket{\Phi^{[1\cdots l]}_{\alpha}} \ket{\Phi^{[(l\!+\!1)\cdots n]}_{\alpha}}.
\label{eq:Schmidt}
\ee
In a generic case $\ch$ grows exponentially with $n$. However, in one-dimensional settings it is sometimes possible to obtain a good approximation to \ket{\Psi} by considering only the first $\ch_{\epsilon}$ terms in (\ref{eq:Schmidt}), with $\ch_{\epsilon} \ll \ch$,
\be
\ket{\Psi} \approx [\sum_{\alpha=1}^{\ch_{\epsilon}} |\lambda^{[l]}_{\alpha}|^2]^{-\frac{1}{2}} \sum_{\alpha=1}^{\ch_\epsilon} \lambda^{[l]}_{\alpha} \ket{\Phi^{[1\cdots l]}_\alpha} \ket{\Phi^{[(l\!+\!1)\cdots n]}_{\alpha}}.
\label{eq:Schmidt2}
\ee
This is due to the following empirical facts, involving the ground state $\ket{\Psi_{gr}}$ and low energy excitations $\ket{\Psi_k}$ of a sufficiently regular, 2-local Hamiltonian $H_n$ (\ref{eq:Hamiltonian}) in one spatial dimension.

{\em Observation 1: Numerical analysis shows that the Schmidt coefficients $\lambda_{\alpha}^{[l]}$ of the ground state $\ket{\Psi_{gr}}$ of $H_n$ decay (roughly) exponentially with $\alpha$,}
\be
\lambda^{[l]}_{\alpha} \sim \exp(-\alpha).
\ee

{\em Observation 2: Numerical analysis shows that during a low-energy evolution, as given by $\ket{\Psi(t)} = \sum_k c_k(t) \ket{\Psi_k}$, the Schmidt coefficients $\lambda_{\alpha}^{[l]}(t)$ also decay (roughly) exponentially with $\alpha$.}

The first observation is at the root of the success of the DMRG in one dimension \cite{trans,Peschel}. The second observation implies that during the time evolution a good approximation to $\ket{\Psi(t)}$ can be obtained by keeping only a small number $\ch_{\epsilon}$ of terms in its Schmidt decomposition, leading to an efficient decomposition ${\cal D}_t$.


{\em Simulation protocol.---} Our aim is to simulate the evolution of the $n$ systems, initially in state $\ket{\Psi_0}$, for a time $T$ according to the Hamiltonian $H_n$ of Eq. (\ref{eq:Hamiltonian}). This is achieved ({\em i}) by constructing decomposition ${\cal D}$ for state $\ket{\Psi_0}$, denoted ${\cal D}_0$, and ({\em ii}) by updating the decomposition ${\cal D}_t$ of the time-evolved state $\ket{\Psi_t}$ for increasing values of a discretized time $t\in\{\delta, 2\delta, \cdots, T\}$, $\delta/T \ll 1$. 

({\em i}) {\em Initialization}. When $\ket{\Psi_0}$ is related to the ground state $\ket{\Psi_{gr}}$ of Hamiltonian $H_n$ by a sufficiently local transformation $Q$ \cite{localQ}, 
\be
\ket{\Psi_0} = Q\ket{\Psi_{gr}},
\ee
${\cal D}_0$ can be obtained from decomposition ${\cal D}_{gr}$ for $\ket{\Psi_{gr}}$ by simulating the action of $Q$ on $\ket{\Psi_{gr}}$ using the present scheme. In turn, ${\cal D}_{gr}$ can be obtained through one of the following three methods: ({\em i1}) by extracting it from the solution of the DMRG method (as detailed elsewhere); ({\em i2}) by considering any product state 
\be
\ket{\Phi_{\otimes}}\equiv \ket{\phi^{[1]}} \otimes \cdots \otimes \ket{\phi^{[n]}},~~~\braket{\Phi_{\otimes}}{\Psi_{gr}}\neq 0,
\ee
for which ${\cal D}$ is always very simple, and by using the present scheme to simulate an evolution in imaginary time $\tau$ according to $H_n$,
\be
\ket{\Psi_{gr}} = \lim_{\tau\rightarrow \infty} \frac{\exp(-H_n\tau) \ket{\Phi_{\otimes}}}{||\exp(-H_n\tau) \ket{\Phi_{\otimes}}||};
\label{eq:imaginary}
\ee
or ({\em i3}) by simulating an adiabatic evolution from some product state $\ket{\Phi'_{\otimes}}$ to $\ket{\Psi_0}$ through a time-dependent Hamiltonian that smoothly interpolates between a local Hamiltonian $H'_n$ such that has $\ket{\Phi_{\otimes}'}$ as its ground state and Hamiltonian $H_n$. 

({\em ii}) {\em Evolution}. For simplicity we assume that $H_n$ does not depend on time \cite{time-dependent}. Then the evolved state reads
\be
\ket{\Psi_T} = \exp(- i H_n T ) \ket{\Psi_0}.
\label{eq:evolution}
\ee
It is convenient to decompose $H_n$ as $H_n=F+G$,
\bea
F &\equiv& \sum_{\mbox{\small even } l} F^{[l]} \equiv \sum_{\mbox{\small even } l} (K_1^{[l]} + K_2^{[l,l\!+\!1]}), \\
G &\equiv& \sum_{\mbox{\small odd } l} G^{[l]} \equiv \sum_{\mbox{\small odd } l} (K_1^{[l]} + K_2^{[l,l\!+\!1]}),
\eea
where $[F^{[l]},F^{[l']}]=0$ ($[G^{[l]},G^{[l']}]=0$) for even (odd) $l,l'$, but possibly $[F,G]\neq 0$. For a small $\delta>0$, the Trotter expansion of order $p$ for $\exp(- i H_n T)$  reads \cite{trotter},
\be
e^{-i(F+G)T} = [e^{-i (F + G) \delta} ]^{T/\delta} 
\approx  [f_p(U_{ F\delta},U_{G\delta })]^{T/\delta},
\label{eq:Trotter}
\ee
where
\bea
U_{ F\delta } \equiv e^{-i F\delta },~~~~~~U_{G\delta } \equiv e^{-i G\delta },
\eea
and
\be
f_1(x,y) = xy,~~~~~~f_2(x,y) = x^{1/2} y x^{1/2},
\ee
for first and second order expansions (see \cite{trotter} for $p=3,4$).
Eq. (\ref{eq:Trotter}) approximates the evolution operator $\exp(- i H_n T)$ by a product of $\O(T/\delta)$ $n$-body transformations $U_{F\delta}$ and $U_{ G\delta}$, which can in turn be expressed as a product of two-body gates $V_2^{[l]}$ and $W_2^{[l]}$,
\bea
U_{F\delta}=\prod_{\mbox{\small even }l} V_2^{[l]}, ~~&& V_2^{[l]} \equiv e^{-iF^{[l]}\delta} ~~\mbox{\small even }l \\
 U_{G\delta}=\prod_{\mbox{\small odd }l} W_2^{[l]}, ~~&& W_2^{[l]} \equiv e^{-iG^{[l]}\delta} ~~\mbox{\small odd }l.
\label{eq:2gates}
\eea
The simulation of the time evolution (\ref{eq:evolution}) is accomplished by iteratively applying gates $U_{F\delta}$ and $U_{G\delta}$ to $\ket{\Psi_0}$ a number ${\cal O} (T/\delta)$ of times, and by updating decomposition ${\cal D}$ at each step. If $\ket{\tilde{\Psi}_t}$ denotes the {\em approximate} evolved state at time $t$, then we have 
\be
\ket{\tilde{\Psi}_{t+\delta}} = f_p(U_{F\delta},U_{G\delta})\ket{\tilde{\Psi}_t},
\ee
where $f_p(U_{F\delta},U_{G\delta})$ consists of a product of ${\cal O}(n)$ two-body gates  $V_2^{[l]}$ and $W_2^{[l]}$, and where we use lemma 2 in \cite{efficient} to update ${\cal D}$ after each of these gates.


{\em Errors and computational cost.--} The main source of errors in the scheme are the truncation (\ref{eq:Schmidt2}) and the Trotter expansion (\ref{eq:Trotter}). We use the fidelity error 
\be
\epsilon(t) \equiv 1 - |\braket{\Psi_t}{\tilde{\Psi}_t}|^2
\ee
to measure how similar the simulated $\ket{\tilde{\Psi}_t}$ and the exact $\ket{\Psi_t}$ are.
The truncation error $\epsilon_1$ incurred in replacing (\ref{eq:Schmidt}) with (\ref{eq:Schmidt2}) reads
\be
\epsilon_1 = \sum_{\alpha=\ch_\epsilon+1}^{\ch} (\lambda^{[l]}_{\alpha})^2.
\label{eq:error1}
\ee
On the other hand, the order-$p$ Trotter expansion neglects corrections $\epsilon_2$ that scale as \cite{error2}
\be
\epsilon_2 \sim \delta^{2p}T^2.
\label{eq:error2}
\ee
Updating ${\cal D}$ after a two-body gate requires ${\cal O}(d^3\ch^3)$ basic operations \cite{efficient}. Gates $U_{F\delta}$ and $U_{G\delta}$ are applied ${\cal O}(T/\delta)$ times and each of them decomposes into about $n$ two-body gates. Therefore ${\cal O}(n(d\ch)^3T/\delta)$ operations are required to apply (\ref{eq:Trotter}) on $\ket{\Psi_0}$. If no truncation takes place, so that the error $\epsilon$ is only due to the Trotter expansion, it follows from (\ref{eq:error2}) that the number of basic operations or {\em computational} time $T_c$ scales as
\be
T_c \sim n(d\ch)^3 \left(\frac{T^{1+p}}{{\epsilon}^{1/2}}\right)^{1/p}.
\label{eq:cost}
\ee


{\em Examples.---} The numerical scheme has been tested in a number of one dimensional settings using matlab code in a pentium IV, both to ($i$) find an approximation to the ground state $\ket{\Psi_{gr}}$ and to ($ii$) simulate time evolutions of local perturbations of $\ket{\Psi_{gr}}$. The next two simple examples aim at illustrating the performance of the method for order $p=2$ Trotter expansions.  

($i$) An approximation to the ground state of a non-critical spin chain, with $n=80$ spins, local dimension $d=2$, $\ch_{\epsilon_1} = 20$ and a sufficiently regular (but otherwise arbitrary) $H_n$ with nearest neighbor interactions can be obtained in half an hour by evolving some product state in imaginary time as in (\ref{eq:imaginary}), until decomposition ${\cal D}_{\tau}$ converges within an accuracy $1-|\braket{\tilde{\Psi}_{\tau}}{\tilde{\Psi}_{\tau + \tau'}}|^2 < 10^{-10}$ \cite{obserbations}.

\begin{figure}
\epsfig{file=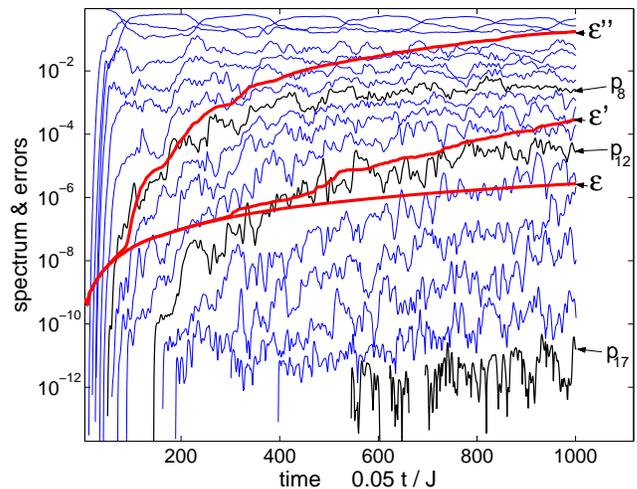,width=8.5cm}
\caption{Propagation of a spin wave in the ferromagnetic chain of Eq. (\ref{eq:ferro}) with $n=30$ spins, $B=J=1$, $T=25$ and $\delta=0.005$. The eigenvalues $p_{\alpha} = |\lambda_{\alpha}^{[15]}|^2$ for the reduced density matrix $\rho_{A}$ for half a chain are plotted as a function of time (thin, irregular, rapidly oscillating lines). Notice the exponential decay of $p_{\alpha}$ in $\alpha$, a seemingly common feature of low-energy dynamics in sufficiently regular, but otherwise arbitrary local models in one dimension. The time scale is such that the spin wave, originating at the left end of the open chain (see state $\ket{\Psi_0}$ after Eq. (\ref{eq:ferro})) runs along the open chain seven times, bouncing backward each time it reaches one extreme spin. The error fidelity $\epsilon(t)$ corresponds to keeping all 17 terms in (\ref{eq:Schmidt}) and grows quadratically in the simulated time $T$ (cf. Eq. (\ref{eq:error2})). Errors $\epsilon'(t)$ and $\epsilon''(t)$ correspond to keeping $\chi_{\epsilon} = 12$ and $\chi_{\epsilon}=8$ terms in Eq. (\ref{eq:Schmidt2}). Truncation errors are of the order of the neglected eigenvalues $p_{\alpha}$. } 
\end{figure}

($ii$) For the spin 1/2 ferromagnetic chain
\be
H_n =  -B\sum_{l=1}^n \sigma_z^{[l]}-  J\sum_{l=1}^{n-1}\vec{\sigma}^{[l]}\cdot \vec{\sigma}^{[l+1]}
\label{eq:ferro}
\ee
the ground state is the product state $\ket{\Psi_{gr}} = \ket{0}^{\otimes n}$ and the exact time evolution of the state $\ket{\Psi_0} = \ket{1}^{\otimes 2}\otimes \ket{0}^{\otimes n-2}$ can be computed efficiently, with a cost that grows only as ${\cal O}(n^2)$, since the dynamics are confined to a subspace of ${\H_2}^{\otimes n}$ of dimension $n(n-1)/2$. In addition, $\ch$ in Eq. (\ref{eq:Schmidt}) is upper bounded by $n/2+2$. Thus we can compare the exact solution $\ket{\Psi_T}$ in (\ref{eq:evolution}) with the approximation $\ket{\tilde{\Psi}_T}$ obtained thorugh simulation. When $\ch_{\epsilon}=n/2+2$, a confirmation of (\ref{eq:cost}) is obtained for different values of $n$, $T$ and $\delta$. For smaller $\ch_{\epsilon}$, truncation errors of the order of the neglected Schimdt terms appear, see Fig. (1). 


In this paper we have shown how to efficiently simulate low energy dynamics in one-dimensional arrays of $n$ quantum systems by using ${\cal O}(n)$ parameters. Since ${\cal O}(\exp(n))$ parameters are required to describe a {\em generic} state of $n$ systems, this result may at first seem contradictory. However, the locality of the interactions and the geometry of the problem make the Hamiltonian $H_n$ highly {\em non-generic}. It is thus conceivable that, correspondingly, the dynamics generated by $H_n$ are constrained to occur in (or can be well approximated by states from) a submanifold of ${\H_d}^{\otimes n}$ with remarkably small dimension. 

Generalizations of the present scheme to Hamiltonians with long-range interactions, to momentum space, to bosonic and fermionic systems, and a specific scheme to address critical systems will be presented elsewhere. These results have also been generalized to {\em slightly correlated} mixed-state dynamics, and in particular to finite temperature quantum many-body dynamics in one dimension.


\medskip

The author thanks Dave Bacon, Ignacio Cirac and David DiVincenzo for valuable advice, and Sergey Bravyi and Miguel Angel Mart\'\i n-Delgado for drawing attention to Refs. \cite{Fannes,Oestlund} after this work was mostly completed.
Support from the US National Science Foundation under Grant No. EIA-0086038 is acknowledged.



\begin{thebibliography}{99}

\bibitem{exact} See for instance S. Albeverio, F. Gesztesy, R. Hoegh-Krohn and H. Holden, {\em Solvable models in Quantum Mechanics} (Springer-Verlag, Texts and Monographs in Physics, 1988, New York); A. G. Ushveridze, {\em Quasi-exactly solvable models in Quantum Mechanics} (Institute of Physics Publishing, London, 1994).

\bibitem{QMC} M. Suzuki, {\em Quantum Monte Carlo Methods in equilibrium and nonequilibrium systems}, Springer Series in solid-state sciences (Springer-Verlag, Berlin, 1986).

\bibitem{DMRG} S. R. White, Phys. Rev. B {\bf 48}, 10345 (1993). S. Rommer and S. \"Ostlund, {\em Density matrix renormalization} Dresden 1998 (Springer, Berlin, 1999).

\bibitem{book} M. A. Nielsen and I. L. Chuang, {\em Quantum Computation and Quantum Information} (Cambridge University Press 2000).

\bibitem{efficient} G. Vidal, Phys. Rev. Lett. {\bf 91}, 147902 (2003), quant-ph/0301063.

\bibitem{Pres} J. Preskill, J. Mod. Opt. {\bf 47}  127-137 (2000), quant-ph/9904022.

\bibitem{trans} G. Vidal, J. I. Latorre, E. Rico, A. Kitaev, Phys. Rev. Lett. {\bf 90} (2003) 227902, quant-ph/0211074.

\bibitem{Nielsen} H. L. Haselgrove, M. A. Nielsen, T. J. Osborne, quant-ph/0308083.

\bibitem{compare} The definition of $\ch$ in the present work, Eq. (\ref{eq:max}), differs from that of Eq. (3) in \cite{efficient}. In the latter the maximization of $\ch_A$ considered all possible blocks $A$ and not just those containing the $m$ first systems. Therefore the condition that $\ch$ be small is here significantly less demanding, as a restriction on entanglement, than the analogous condition in \cite{efficient}.

\bibitem{Fannes} M. Fannes, B. Nachtergaele and R. F. Werner, Comm. Math. Phys. {\bf 144}, 3 (1992), pp. 443-490; J. Funct. Anal. {\bf 120}, 2 (1994), pp. 511-534.

\bibitem{Affleck} I. Affleck, T. Kennedy, E. H. Lieb and H. Tasaki, Phys. Rev. Lett. {\bf 59}, 799 (1987); Commun. Math. Phys. {\bf 155}, 477 (1988).

\bibitem{Oestlund} S. \"Ostlund and S. Rommer, Phys. Rev. Lett. {\bf 75}, 19 (1995), pp. 3537. 

\bibitem{Schmidt} E. Schmidt, Math Ann. {\bf 63}, 433 (1907); A. Ekert and P. L. Knight, Am. J. Phys. {\bf 63}, 415 (1995); A. Peres, {\em Quantum theory: concepts and methods} (Kluwer Academic Publishers, Dordrecht, 1995).

\bibitem{Peschel} see I. Peschel, J. Phys. A: Math. Gen. {\bf 36}, L205 (2003) and references therein.

\bibitem{localQ} Operator $Q=Q_p \cdots Q_2 Q_1$ could be the product of a small number $p$ of operators $Q_m$, where each $Q_m$ is at most a two-body transformation, and such that for $m=1,\cdots, p$, vector $Q_m\cdots Q_1\ket{\Psi_0}\in {H_d}^{\otimes n}$ can be efficiently described using ${\cal D}$. Examples are given by a product of operators $\sigma^{+}\equiv (\sigma_x - i \sigma_y)/2$ acting on several sites of a spin chain, or a product of (real space) creation operators $a^{\dagger}$ in a 1D bosonic or fermionic system. 

\bibitem{time-dependent} If $H_n$ depends on time, we consider time intervals $\Delta T$ such that $H_n$ is approximately constant between $t$ and $t+\Delta T$ and to each interval we apply the same reasoning as in the time-independent case.

\bibitem{trotter} M. Suzuki, Phys. Lett. A {\bf 146}, 6, (1990) 319-323; J. Math. Phys. {\bf 32}, 2 (1991) 400-407. For third and forth order expansions see also A. T. Sornborger and E. D. Stewart, quant-ph/9809009.

\bibitem{error2} The error $\zeta \equiv tr|e^{-iH_nT}- {f_p}^{T/\delta}|$ (cf. Eqs. (\ref{eq:evolution}) and (\ref{eq:Trotter})) scales as $T/\delta$ times $\delta^p$ \cite{trotter} for an order $p$ Trotter expansion. The approximated state $\ket{\tilde{\Psi}_T}$ is, up to a phase, $\ket{\tilde{\Psi}_T} = (1-{\cal O} (\zeta^2)) \ket{\Psi_T}+ {\cal O}(\zeta)\ket{\Psi^{\perp}(T)}$, so that $\epsilon \equiv 1 - |\braket{\Psi}{\tilde{\Psi}}|^2$ scales as $\zeta^2 \sim \delta^{2p}T^2$.

\bibitem{obserbations} We observe that convergence is achieved significantly faster for non-critical chains than for critical chains. This was to be expected, since in a critical chain the gap in the spectrum of $H_n$ is small (and vanishes in the limit of large $n$). In addition, the particular scaling of entanglement at a quantum phase transition \cite{trans} implies that the number $\ch_{\epsilon}$ of Schmidt terms required at a critical point is larger than that required in a localized phase in order to obtain the same truncation error $\epsilon_1$. Finally, in some limiting cases corresponding to highly delocalized (and thus very entangled) phases, a small $\ch_{\epsilon}$ does not lead to a good approximation to $\ket{\Psi_{gr}}$ because $\lambda^{[l]}_{\alpha}$ happens to decay slowly with $\alpha$.

\end{thebibliography}
\end{document}